\documentclass[fleqn]{bmcart}      
\usepackage[numbers,sort&compress]{natbib}
\usepackage{graphicx}
\usepackage{amsmath}
\usepackage{amssymb}
\usepackage{xcolor}
\usepackage{enumitem}
\usepackage[utf8]{inputenc} 

\startlocaldefs
\newcommand{\rg}{r_{\rm g}}
\newcommand{\tunit}{r_{\rm g}/c}
\newcommand{\msun}{\mathrm{M}_{\odot}}
\endlocaldefs

\begin{document}

\begin{frontmatter}
 
 \begin{fmbox}
  \dochead{Research}
  
  
  \title{Observing Supermassive Black Holes In  Virtual Reality}

  \author[
  addressref={aff1},
  corref={aff1},
  noteref={},
  email={j.davelaar@astro.ru.nl}
  ]{\inits{JD}\fnm{Jordy} \snm{Davelaar}}
  \author[
  addressref={aff1},
  email={}
  ]{\inits{TB}\fnm{Thomas} \snm{Bronzwaer}}
  \author[
  addressref={aff1},
  email={}
  ]{\inits{DK}\fnm{Daniel} \snm{Kok}}
  \author[
  addressref={aff2,aff3},
  email={}
  ]{\inits{ZY}\fnm{Ziri} \snm{Younsi}}
  \author[
  addressref={aff1},
  noteref={},
  email={}
  ]{\inits{MM}\fnm{Monika} \snm{Mościbrodzka}}
  \author[
  addressref={aff1},
  email={}
  ]{\inits{HF}\fnm{Heino} \snm{Falcke}}


  \address[id=aff1]{
   \orgname{Department of Astrophysics/IMAPP, Radboud University Nijmegen},
   \street{P.O. Box 9010},
   \postcode{65008}
   \city{Nijmegen},
   \cny{The Netherlands}
  }
  \address[id=aff2]{
   \orgname{Institute for Theoretical Physics},
   \street{Max-von-Laue-Str. 1},
   \postcode{60438}
   \city{Frankfurt am Main},
   \cny{Germany}
  }
  
  \address[id=aff3]{
   \orgname{Mullard Space Science Laboratory, University College London},
   \street{Holmbury St. Mary, Dorking},
   \postcode{RH5 6NT}
   \city{Surrey},
   \cny{United Kingdom}
  }
  
  
  \begin{artnotes}
  \end{artnotes}
  
 \end{fmbox}
 
 
 \begin{abstractbox}
  
  \begin{abstract} 
   We present a full $360^{\circ}$ (i.e., $4\pi$ steradian) general-relativistic ray-tracing and
   radiative transfer calculations of accreting supermassive black holes.
   We perform state-of-the-art three-dimensional general-relativistic magnetohydrodynamical
   simulations using the \texttt{BHAC} code, subsequently post-processing this data with the radiative
   transfer code \texttt{RAPTOR}.
   All relativistic and general-relativistic effects, such as Doppler boosting and gravitational redshift, as well
   as geometrical effects due to the local gravitational field and the observer's changing position and state of motion,
   are therefore calculated self-consistently.
   Synthetic images at four astronomically-relevant observing frequencies are generated from the perspective of an observer with a full $360^{\circ}$
   view inside the accretion flow, who is advected with the flow as it evolves.
   As an example we calculated images based on recent best-fit models of observations of Sagittarius A*. These images are combined to
   generate a complete $360^{\circ}$ Virtual Reality movie of the surrounding environment of the black hole and its event horizon.
   Our approach also enables the calculation of the local luminosity received at a given fluid element in the accretion flow, providing important
   applications in, e.g., radiation feedback calculations onto black hole accretion flows.
   In addition to scientific applications, the $360^{\circ}$ Virtual Reality movies we present also represent a new medium through
   which to interactively communicate black hole physics to a wider audience, serving as a powerful educational tool.
  \end{abstract}
  
  
  
  
 \end{abstractbox}
 %
 
\end{frontmatter}



\section*{Main text}

\section{Introduction}
Active Galactic Nuclei (AGN) are strong sources of electromagnetic radiation from the radio up to $\gamma$-rays.
Their source properties can be explained in terms of a galaxy hosting an accreting supermassive black hole
(SMBH) in its core.
The Milky Way also harbours a candidate SMBH, Sagittarius A* (Sgr~A*), which is subject to intensive
Very-Long-Baseline Interferometric (VLBI) studies
\cite{krichbaum1998,bower2004,shen2005,doeleman,bower2014,brinkerink2016}.
Sgr~A* is one of the primary targets of the Event Horizon Telescope Collaboration (EHTC), which aims to image for the very first time the ``shadow" of
a black hole \cite{Goddi2017}.
Theoretical calculations predict this shadow to manifest as a darkening of the inner accretion flow image
anticipated to be observed due to the presence of a black hole event horizon, representing the region within which
no radiation can escape \cite{grenzebach2016,Goddi2017,younsi2016}.
The apparent size on the sky of this shadow is constrained by Einstein's General Theory of Relativity (GR)
\cite{bardeen1973,cunningham1973,luminet,viergutz,falckemelia,johannsen2010,johannsen2013,younsi2016},
and observational measurements of the black hole shadow size and shape can in principle provide a strong test of
the validity of GR in the strong-field regime \cite{johannsen2010,Abdujabbarov2015,younsi2016,Goddi2017}.

The theoretical aspects of the observational study of Sgr~A* require the generation of general-relativistic magnetohydrodynamical (GRMHD) simulation data of
the accretion flow onto a black hole, which is subsequently used to calculate synthetic observational data for
physically-motivated plasma models which can be compared to actual observational data.
In the past, synthetic observational data was generated by ray-tracing radiative transfer codes
which calculate the emission originating from the accreting black hole and measured by a far away observer by solving
the equations of radiative transfer along geodesics, i.e., the paths of photons (or
particles) as they propagate around the black hole in either static spacetimes \citep[e.g.][]
{broderick2006,noble2007,dexter2009,shcherbakov2011,vincent2011,younsi2012,chan2013,younsi2015,
 dexter2016,schnittman2016,chan2017,moscibrodzka2017,bronzwaer2017} or dynamical spacetimes \cite{kelly2017,schnittman2018}.

These models vary only in the dynamics of the black hole accretion flow, with the observer remaining stationary
through the calculations.
In this work, we consider the most general case of an observer who can
vary arbitrarily in both their position (with respect to the black hole) and their state of motion.
In particular, the observer is chosen to follow the flow of the accreting plasma in a physically-meaningful
manner through advection, and therefore all dynamical effects introduced by the motion of the observer around the
black hole are also correctly included in the imaging calculation.

With recent developments in Graphical Processor Units (GPUs) and Virtual Reality (VR) rendering, it has become
possible to visualise these astrophysical objects at high resolutions in a $360^{\circ}$ (i.e., $4\pi$ steradian) format that covers the entire celestial sphere of an observer, enabling the study of the surroundings of an accreting black hole from within the accretion flow itself. Virtual Reality is a broad concept that encompasses different techniques, such as immersive visualisation, stereographic rendering, and interactive visualisations. In this work, we explore the first of these three, by rendering the full celestial sphere of the observer along a trajectory. The viewer can then look in any direction during the animation; this is also known as $360^{\circ}$ VR. Another important feature of VR, stereographic rendering, presents different images to each eye, so that the viewer experiences stereoscopic depth. For our application, however, this technique is not relevant, since the physical distance between the eyes of the observer is much smaller than the typical length scale of a supermassive black hole (which is $ 6.645\times10^{11}$ cm for Sagittarius A*), and therefore we would not see any depth in the image (just as we do not see stereoscopic depth when looking at the Moon). Interactive visualisations, where the viewer also has the freedom to change his or her position, would require real-time rendering of the environment, which is beyond the reach of current computational resources.

Our new way of visualising black holes enables the study of accretion from the point of view of an observer
close to the black hole event horizon, with the freedom to image in all directions, as opposed to the perspective of
an observer far away from the source with a fixed position and narrow field of view.
In the case of a distant observer, the source appears projected onto the celestial sphere
(thus appearing two-dimensional).
Since one cannot easily distinguish three-dimensional structures within the accretion flow, placing the observer
inside the flow itself opens a new window in understanding the geometrical structure and dynamical properties of
such systems.
Several researchers have previously considered an observer moving around, or falling into a black hole, e.g.,
\begin{enumerate}[label={(\arabic*)}]
 \item falling through the event horizon as illustrated through the gravitational lensing distortions of different regions
       (e.g., the ergo-region and event horizon), represented as chequerboard patterns projected onto an observer's
       image plane \cite{madore},
 \item a flight through a simulation of a non-rotating black hole \cite{hamilton},
 \item a flight through an accretion disk of a black hole using an observer with a narrow field of view camera
       \cite{luminet2011},
 \item a $360^{\circ}$ VR movie of an observer falling into a black hole surrounded by vacuum with illumination provided exclusively by background starlight, i.e., without an accretion flow \cite{younsi},
 \item a $360^{\circ}$ VR movie of a hotspot orbiting a SMBH \cite{monikaVR}, and
 \item a $360^{\circ}$ VR movie of an N-body/hydrodynamical simulation of the central parsec of the Galactic center \cite{russelVR}.
\end{enumerate}

In this study, we consider a self-consistent three-dimensional GRMHD simulation of the accretion
flow onto a spinning (Kerr) black hole, determining its time evolution and what an observer would see in full
$360^{\circ}$ VR as they move through the dynamically evolving flow.
To image accreting black holes in VR, we use the general-relativistic radiative-transfer (GRRT)
code \texttt{RAPTOR} \citep{bronzwaer2017}.
The code incorporates all important general-relativistic effects, such as Doppler boosting and gravitational lensing
in curved spacetimes, and can be compiled and run on both Central Processing units (CPU's) and GPU's by using {NVIDIA's}
\texttt{OpenACC} framework.

In this work, we investigate the environment of accreting black holes from within the accretion flow itself with a
virtual camera.
As an example astrophysical case we model the supermassive black hole Sgr~A*, although
the methods presented in this work are generally applicable to any black hole as long as the radiation
field's feedback onto the accreting plasma has a negligible effect on the plasma's magnetohydrodynamical
properties, which is the case for Low Luminosity AGNs or low/hard state X-ray binaries.

The trajectory of this camera consists of two phases: a hovering trajectory, where the observer moves with a pre-defined velocity, and a particle trajectory, where the observer's instantaneous velocity is given by a trajectory of a tracer particle computed with a seperate axisymetric GRMHD simulation. The tracer particle follows the local plasma velocity (specifically, it is obtained by interpolating the plasma velocity of the GRMHD simulation cells to the camera's location).

We present a $360^{\circ}$ VR simulation of Sgr~A*, demonstrating the applications of VR for
studying not just accreting black holes but also for education,
public outreach and data visualisation and interpretation amongst the wider scientific community.
In section \ref{sec:methods} we describe the camera setup, present several black hole shadow lensing
tests, describe the camera trajectories and outline the radiative transfer calculation.
In section \ref{sec:results} we present our $360^{\circ}$ VR movie of an accreting black hole.
In section \ref{sec:discconlc} we discuss our results and outlook.

\section{\label{sec:methods} Methods}
In this section, we introduce the virtual camera setup, present black hole shadow vacuum lensing tests using both
stationary and free-falling observers at different radial positions, discuss the different camera trajectories used in the VR movie shown later in this article and introduce the GRMHD plasma model that is used as an input for the geometry of the accretion flow onto the black hole.

The original {\tt RAPTOR} code \cite{bronzwaer2017} initialises rays (i.e., photon geodesics) using impact
parameters determined form coordinate locations on the observer's image plane \cite{bardeen1972}.
This method is not suitable for VR since it only applies to distant observers where geometrical distortions in the image which arise from the strong gravitational field (i.e., spacetime curvature) of the black hole are negligible.
To generate full $360^\circ$ images as seen by an observer close to the black hole, we have extended the
procedure of \cite{noble2007} to use an orthonormal tetrad basis for the construction of initial photon wave
vectors, distributing them uniformly as a function of $\theta\in[0,\pi]$ and $\phi\in[0,2\pi]$ over a unit sphere.

The advantage of this approach is that all geometrical, relativistic, and general- relativistic effects on
the observed emission are naturally and self-consistently folded into the imaging calculation,
providing a complete and physically-accurate depiction of what would really be seen from an observer's
perspective.

The first step in building the tetrad basis is using a set of \emph{trial vectors} (specifically, 4-vectors),
$t_{(a)}^{\mu}$, to find the \emph{tetrad basis vectors}, $e_{(a)}^{\mu}$.
Herein, parenthesised lowercase Roman letters correspond to tetrad frame indices
while Greek letters correspond to coordinate frame indices.
Unless stated otherwise, all indices are taken to vary over $0$--$3$, with $0$ denoting the temporal component
and $1$--$3$ denoting the spatial components of a given 4-vector.
Given a set of $\left\{\theta,\phi\right\}$ pairs (typically on a uniform grid), the corresponding wave
vector components in the tetrad frame, $k^{(a)}$, are given by:
\begin{align}
 k^{(0)} = & +1 \,, \label{eqn:wavevector1}                       \\
 k^{(1)} = & -\cos(\phi) \cos(\theta) \,, \label{eqn:wavevector2} \\
 k^{(2)} = & -\sin(\theta) \,, \label{eqn:wavevector3}            \\
 k^{(3)} = & -\sin(\phi)\cos(\theta) \,, \label{eqn:wavevector4}  
\end{align}
where it is trivial to verify that this wave vector satisfies $k_{(a)} k^{(a)} = 0$,
as expected for null geodesics.

In order to determine the wave vector defined in eqs.~\eqref{eqn:wavevector1}--\eqref{eqn:wavevector4} in the
coordinate frame, $k^{\alpha}$, it is necessary to first construct the tetrad vectors explicitly.
The first trial vector we use is the four-velocity of the observer,
$t_{(0)}^{\mu} = u^\mu_{\rm obs}$.
This vector is, by virtue of sensible initial conditions and preservation of the norm
during integration, normalised.
Using the four-velocity as an initial trial vector also ensures that Doppler effects due to the motion of the
camera is included correctly.
It is then possible to build a set of orthonormal basis vectors $e_{(a)}^{\mu}$
by using the Gram-Schmidt orthonormalisation procedure.
The required trial vectors for this procedure are given by:
\begin{eqnarray}
 t_{(1)}^{\mu} &=& (0, -1, 0, 0) \,, \\
 t_{(2)}^{\mu} &=& (0,  0, 1, 0) \,, \\
 t_{(3)}^{\mu} &=& (0, 0 , 0, 1) \,.
\end{eqnarray}
This set of trial vectors is chosen such that the observer always looks towards the black hole in a right-handed
basis.
Any other initialisation, e.g., along with the velocity vector, could cause discomfort when used in VR due to high
azimuthal velocities.
The wave vector may now be found by taking the inner product of the tetrad basis
vectors and the wave vector in the observer's frame as:
\begin{equation}
 k^\mu = e_{(a)}^\mu k^{(a)} \,.
\end{equation}
The observer's camera is then initialised at a position $X_{\rm cam}^\mu$ and uniformly-spaced rays are launched
in all directions from this point.
This method is fully covariant and is therefore valid in any coordinate system.
\subsection{Black holes and gravitational lensing}
In this work, we adopt geometrical units, $G=M=c=1$, such that length and time scales are dimensionless.
Hereafter $M$ denotes the black hole mass, and setting $M=1$ is equivalent to rescaling the length scale
to units of the gravitational radius, $r_{\rm g} := GM/c^2$, and the time scale to units of $r_{\rm g}/c = GM/c^3$.
To rescale lengths and times to physical units, one simply scales $r_{\rm g}$ and  $r_{\rm g}/c$ using
the appropriate black hole mass.
For Sgr~A* these scalings are given by $r_{\rm g} = 5.906 \times 10^{11}$ cm and $r_{\rm g}/c = 19.7$ seconds,
respectively.

The line element in GR determines the separation between events in space-time, and is defined as:
\begin{equation}
 ds^2 = g_{\mu\nu} \ \! dx^\mu dx^\nu \,,
\end{equation}
where $g_{\mu\nu}$ is the metric tensor and $dx^\mu$ an infinitesimal displacement vector.
The metric is a geometrical object that contains all the information concerning the space-time under consideration
(in this study a rotating Kerr black hole) and is used to raise and lower tensor indices, e.g.,
$g_{\alpha\mu}A^{\mu\nu_{1}\nu_{2}\ldots\nu_{\rm n}} =
A^{\phantom{\alpha}\nu_{1}\nu_{2}\ldots\nu_{\rm n}}_\alpha$, where the Einstein summation convention is
implicitly assumed.
The line element for a rotating black hole is given by the Kerr metric \cite{Kerr1963}, which is written in
Boyer-Lindquist coordinates $x^{\mu}=\left( t,r,\theta,\phi \right)$ as:
\begin{eqnarray}
 ds^2=&-&\left(1-\frac{2r}{\Sigma}\right) dt^2 -\frac{4ar\sin^2\theta}{\Sigma}dt \, d\phi
 +\frac{\Sigma}{\Delta}dr^2 + \Sigma d\theta^2 \nonumber \\
 &+&\left(r^2+a^2+ \frac{2ra^2\sin^2\theta}{\Sigma}\right)\sin^2\theta \, d\phi^2 \,, \label{KerrMetricBL}
\end{eqnarray}
where
\begin{eqnarray}
 \Delta &:=& r^2 - 2r + a^2, \\
 \Sigma&:=&r^2 +a^2 \cos^2\theta,
\end{eqnarray}
and $a$ is the dimensionless spin parameter of the black hole.

In the above form, the Kerr metric has a coordinate singularity at the outer (and inner) event horizon,
which presents difficulties for both the numerical GRMHD evolution and the GRRT calculations.
This also prohibits the observer's camera from passing smoothly through this region.
To avoid this we transform \eqref{KerrMetricBL} from $x^{\mu}$ into horizon-penetrating Kerr-Schild
coordinates $\tilde{x}^{\mu}=\left(\tilde{t}, \tilde{r}, \tilde{\theta}, \tilde{\phi} \right)$ as:

\begin{equation}
 \tilde{t} = t + \ln\Delta + 2 \mathcal{R} \,, \quad
 \tilde{r} = r \,, \quad
 \tilde{\theta} = \theta \,, \quad
 \tilde{\phi} = \phi + a \mathcal{R} \,,
\end{equation}
where
\begin{equation}
 \mathcal{R} \equiv \frac{1}{r_{\rm out}-r_{\rm in}}\ln\left( \frac{r-r_{\rm out}}{r-r_{\rm in}} \right) \,. \label{R_horizon}
\end{equation}
In eq.~\eqref{R_horizon} the outer horizon is given by $r_{\rm out} \equiv 1 + \sqrt{1-a^{2}}$, and the inner horizon
by $r_{\rm in} \equiv 1 - \sqrt{1-a^{2}}$.
Hereafter the coordinate system employed in this study is the modified Kerr-Schild (MKS) system,
denoted by $X^{\mu}$, which is related to the aforementioned Kerr-Schild coordinates, $\tilde{x}^{\mu}$, as:
\begin{equation}
 X^{0} = \tilde{t}\,, \quad X^{1} = \ln \tilde{r}\,, \quad X^{2} = \tilde{\theta}/\pi\,, \quad X^{3} = \tilde{\phi} \,.
\end{equation}
To visualise the effect of a moving camera compared to a stationary camera, we calculate light rays
originating from both a stationary observer and a free-falling observer.
This calculation is performed at two different positions, which in MKS coordinates are given by:
\begin{equation}
 X^\mu_1 = ( 0, \, \ln 10,\, 0 , \,0) \! \qquad {\rm and} \qquad \! X^\mu_2 = ( 0, \, \ln 3, \,0 , \,0) \,.
\end{equation}
Consequently, the observer positions $1$ and $2$ correspond to radial distances of $10~\rg$ and $3~\rg$,
respectively.
An observer at rest has a four-velocity
\begin{equation}
 u^\mu_{0} = \left( \alpha, \, 0, \, 0, \, 0 \right) \,,
\end{equation}
where $\alpha:=\left(-g^{tt}\right)^{-1/2}$ is the lapse function.
At the positions $X^\mu_{1}$ and $X^{\mu}_{2}$ the free-falling observer has the following corresponding
four-velocity components:
\begin{equation}
 u^\mu_{1} = ( 1.10, \, -0.029, \, 0, \, -0.0011) \qquad {\rm and} \qquad
 u^\mu_{2} = ( 1.34, \, -0.26, \, 0, \, -0.034) \,.
\end{equation}
The free-falling velocities were obtained by numerically integrating the geodesic equation for a free-falling
massive particle.

To visualise the effect of the observer's motion on the observed field of view, we place a sphere around
both the observer and the black hole, which is centred on the black hole.
This is what we subsequently refer to as the ``celestial sphere".
The black hole spin is taken to be $a=0.9375$, the exact value of the spin parameter for Sgr~A* is unknown, the chosen value was the best fit of a parameter survey \cite{monika2009}.
The observer is positioned in the equatorial plane of the black hole (i.e., $\theta=90^{\circ}$),
where the effects of gravitational lensing are most significant and asymmetry in the shadow shape
due to the rotational frame dragging arising from the spin of the black hole is most pronounced.

\begin{figure}[htbp]
 \centering
 \includegraphics[width=0.9\textwidth]{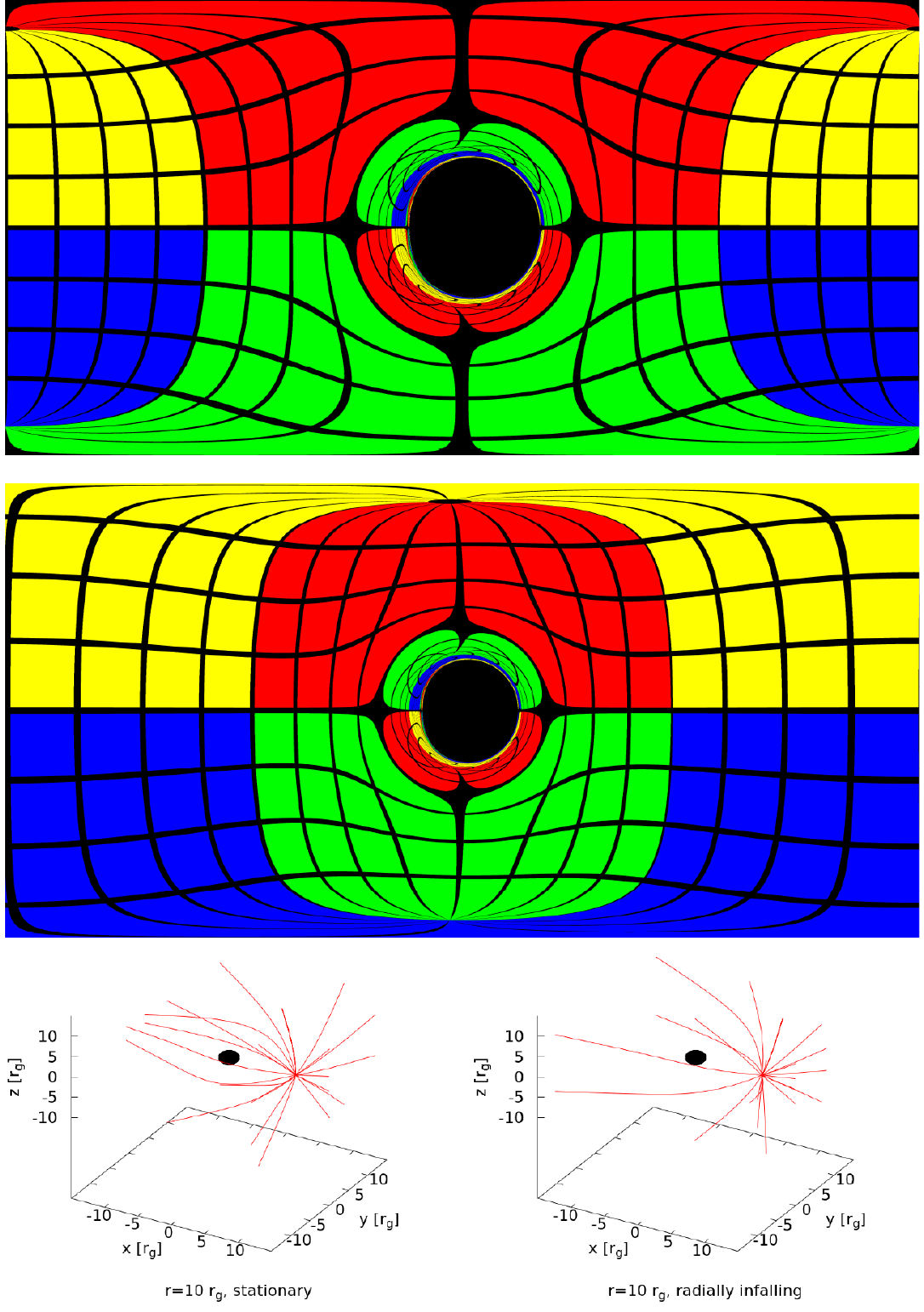}
 \caption[Celestial sphere and black-hole shadow images for an observer located at $r=10~\rg$.]{ Celestial sphere and black hole shadow images for an observer located at $r=10~\rg$.
  Top panel: celestial sphere and shadow image as seen by a {\it stationary} observer. The different colors represent different quadrants of the sky, yellow and blue being behind the observer, while red and green are in front of the observer. The black lines represent lines of constant longitude and lattitude while the black, circular region in the center is the black-hole shadow.
  Middle panel: as top panel, but seen by a {\it radially in-falling} observer.
  Bottom-left panel: photons originating from a stationary observer's camera, as used to generate the top panel.
  Bottom-right panel: photons originating from a radially in-falling observer's camera, as used to generate the
  middle panel.
  The black hole event horizon is shown as the black region in both bottom panels.
  The shadow sizes are similar in both panels, but differences are clearly visible.
  See corresponding text for further discussion.
 }
 \label{fig:celestial_1}
\end{figure}

\begin{figure}[htbp]
 \centering
 \includegraphics[width=0.9\textwidth]{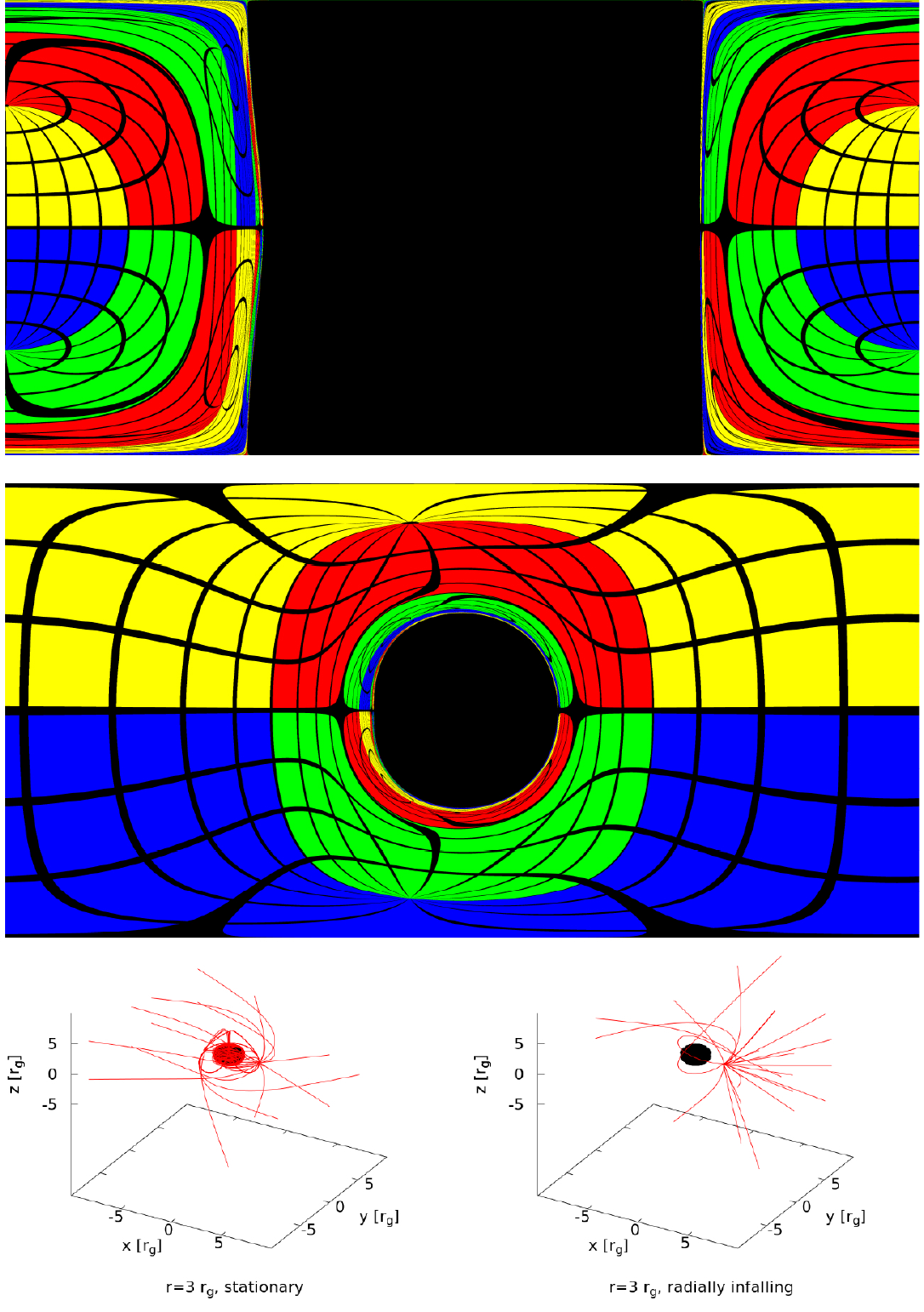}
 \caption[Celestial sphere and black hole shadow images for an observer located at $r=3~\rg$.]{As in Fig.~\ref{fig:celestial_1}, now with the observer located at $r=3~\rg$.
  Differences between the shadow size and shape as seen by the two observers are now significant.
  See corresponding text for further discussion.
 }
 \label{fig:celestial_2}
\end{figure}

Each quadrant of the celestial sphere is then painted with a distinct colour and lines of constant longitude
and latitude are included to aid in the interpretation of the angular size and distortion of the resulting images. The celestial sphere in Minkowski spacetime,
where we used cartesian coordinates to integrate the geodesics, as seen by an observer positioned at $10 ~\rg$ can be seen in Figure \ref{fig:celestial_0}.
The number of coloured patches in the $\theta$ and $\phi$ directions is
$\left(n_{\theta},n_{\phi}\right)=\left(8,16\right)$.
Therefore, excluding the black lines of constant latitude and longitude
(both $1.08^{\circ}$ in width), each coloured patch subtends an angle of $22.5^{\circ}$ in both
directions.
We also calculated 25 light rays for each of these observers, distributing them equally over $\left(\theta,\phi\right)$
in the frame of the observer (see bottom rows of Figs.~\ref{fig:celestial_1} \& \ref{fig:celestial_2}) in order
to interpret the geometrical lensing structure of the images in terms of their constituent light rays.

Figure \ref{fig:celestial_1} presents black hole shadow images and background lensing patterns for the Kerr
black hole as seen by both a stationary observer (top panel) and a radially infalling observer (middle panel)
located at a distance of $10~\rg$.
The angular size of the shadow is larger for the stationary observer.
This observer, being in an inertial frame, is essentially accelerating such that the local gravitational
acceleration of the black hole is precisely counteracted by the acceleration of their reference frame.
This gives rise to a force on the observer directed away from the black hole itself, reducing the angular
momentum of photons oriented towards the black hole (seen as the innermost four rays being bent
around the horizon), effectively increasing the black hole's capture cross-section and producing a larger shadow.
Strong gravitational lensing of the image due to the presence of the compact mass of the black hole is
evident in the warping of the grid lines.

In Figure \ref{fig:celestial_2} the observers are now placed at $3~\rg$, i.e., very close to the black hole.
For the stationary observer, all photons within a field of view centred on the black hole of $>180^{\circ}$ in the
horizontal direction and over the entire vertical direction, are captured by the black hole.
Such an observer looking at the black hole would see nothing but the darkness of the black hole shadow
in all directions.
This is clear in the corresponding bottom-left plot of photon trajectories.
As the observer approaches the event horizon the entire celestial sphere begins to focus into an ever shrinking
point adjacent to the observer.
For the infalling observer, the lensed image is far less extreme.
Whilst the shadow presents a larger size in the observer's field of view, this is mostly geometrical, i.e., due
to the observer's proximity to the black hole.
There is also visible magnification of regions of the celestial sphere behind the observer.
These results clearly follow from the photon trajectories in the bottom-right panel.

In all images of the shadow, repeated patches of decreasingly small area and identical colours are visible.
In particular, multiple blue and yellow patches whose photons begin from behind the observer are visible
near the shadow.
These are a consequence of rays which perform one or more orbits of the black hole before reaching the
observer, thereby appearing to originate from in front of the observer.

\begin{figure}[htbp]
 \centering
 \includegraphics[width=0.9\textwidth]{f0.pdf}
 \caption[Celestial sphere in Minkowski spacetime for an observer located at $r=10~\rg$.]{Celestial sphere in Minkowksi spacetime for an observer at $r=10~\rg$. The different colors represent different quadrants of the sky, with yellow and blue being behind the observer, while red and green are in front of the observer. The black lines represent lines of constant longitude and lattitude.
 }
 \label{fig:celestial_0}
\end{figure}

\subsection{Camera trajectories}
As described in Section 1, we consider two distinct phases for the camera trajectory.
The first phase assumes a hovering observer positioned either at a fixed point or on a hovering trajectory
around the black hole (i.e., the camera's motion is unaffected by the plasma motion and is effectively in an inertial frame).
For the second phase of the trajectory, the observer's four-velocity is determined from an axisymmetric
GRMHD simulation which includes tracer particles that follow the local plasma velocity. The choice to perform a separate tracer-particle simulation that is axisymmetric, in contrast to the 3D plasma simulation, was made to omit turbulent features in the $\phi$ direction which can be nauseating to watch in VR environments. This makes the movie scientifically less accurate, but is necessary to prevent viewers from experiencing motion sickness. Since the methods presented in this paper are not dependent on the dimensionality of the tracer particle simulation, they can be used for full 3D tracer particle simulations as well.
In the following subsections, these two camera trajectories are described in detail.

\subsubsection{Hovering trajectory}
In the first phase of the trajectory, the observer starts in a vacuum, with only the light from
the distant background stars being considered in the calculation.
The observer is initially at a radius of $400~\rg$ and moves inward to $40~\rg$.
After this, the observer rotates around the black hole, which we term the ``initialisation scene'', and comprises $1600$ frames.
Each frame is separated by a time interval of $1~\tunit$.
The first phase of the movie, which includes the time-evolving accretion flow, consists of $2000$ frames from the
perspective of an observer at a radius of $40~\rg$ and an inclination of $60^\circ$ with respect to the
spin axis of the black hole.
We refer to this first phase as ``Scene 1''.
We then subsequently rotate around the black hole whilst simultaneously moving inward to a radius of $20~\rg$
over a span of $1000$ frames, which we refer to as ``Scene 2''.
Within Scene 2, after the first $500$ frames the observer then starts to decelerate until stationary once more.

\subsubsection{\label{particleflow} Particle trajectory}

For the second phase of the trajectory, the observer moves along a path that is calculated from an axisymmetric
GRMHD simulation which includes tracer particles.
The tracer particles act like test masses: their velocity is found by interpolating the local plasma
four-velocity (which is stored in a grid-based data structure) to the position of the particle.
A first-order Euler integration scheme is then employed to update the position of each particle.
For the camera, we are concerned with particles which are initially located within the accretion disk,
begin to accrete towards the black hole, and then subsequently leave the simulation domain via the jet.
To identify particles which satisfy all of these conditions we create a large sample of particle trajectories.
The number of injected particles, ${\mathcal{N}}_{\rm inj}$, within a grid cell with index $\{i,j\}$ is set by
two parameters: the plasma density, $\rho$, of the bounding cell, and the total mass, $M_{\rm tot}$, within
the simulation domain.
The number of injected particles is then calculated as
\begin{equation}
 {\mathcal{N}}_{\rm inj}\left(i,j\right) = N_{\rm tot} \left( \frac{\rho\left(i,j\right) V_{\rm cell}}{M_{\rm tot}} \right) \,,
\end{equation}
where the weight factor ensures that only a predefined number of particles, $N_{\rm tot}$,
after appropriate weighting, are then injected into a given simulation cell of volume
$V_{\rm cell} = \sqrt{-g} \ \! dx^1 dx^2 dx^3$, where $g$ is the determinant of the metric tensor.
The code then randomly distributes these particles inside the simulation cell.
\begin{figure}[htbp]
 \centering
 \includegraphics[width=0.95\textwidth]{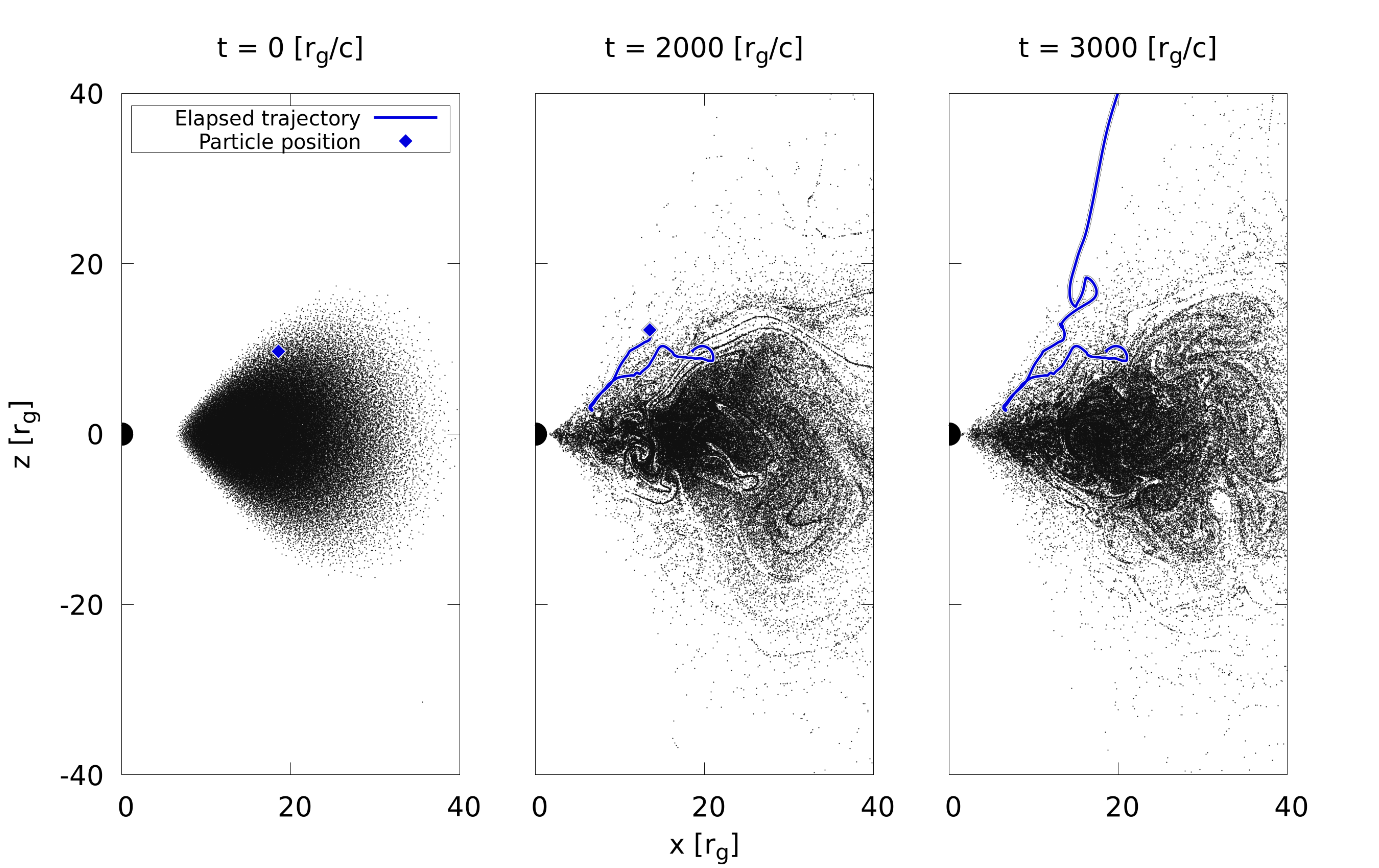}
 \caption[Snapshots of the tracer particles in the advection simulation]{Left panel: initial distribution of particles inside the initial torus.
  Middle panel: snapshot of the advection {\tt HARM2D} simulation at $t=2000~\tunit$.
  Right panel: later snapshot at $t=4000~\tunit$.
  The two times correspond to the advection simulation time, i.e., frames $4600$--$7600$ in the resulting movie.
  The blue square represents the initial position of the tracer particle used for the camera.
 The blue curve shows the trajectory corresponding to this tracer particle.}
 \label{fig:particles}
\end{figure}
The particles are initially in Keplerian orbits and co-rotate with the accretion disk.
The disk then quickly becomes turbulent due to the growth of the magneto-rotational instability (MRI).
As the particles are advected with the flow they can be classified into three different types:
\begin{enumerate}[label={(\arabic*)}]
 \item {\it accreted particles} which leave the simulation at the inner radius (i.e., plunge into the event horizon) and
       remain gravitationally bound,
 \item {\it wind particles} which become gravitationally unbound, travel through weakly magnetised regions and
       then exit the simulation at the outer boundary,
 \item {\it accelerated jet particles} which are similar to wind particles but additionally undergo rapid
       acceleration within the highly-magnetised jet sheath.
\end{enumerate}
To discriminate between these three types of particle, several key hydrodynamical and magnetohydrodynamical
criteria are examined.
The first criterion is that the hydrodynamical Bernoulli parameter of the particle satisfies
${\rm Bern} = - h u_{\rm t} > 1.02$, where $h$ is the (specific) enthalpy of the accretion flow and
$u_{\rm t}$ is the covariant time component of the four-velocity.
When this condition is satisfied the particle is, by definition, unbound. The boundary transition between bound and unbound happens at ${\rm Bern} = - h u_{\rm t} > 1.00$,
but we take a slightly larger value to select the part of the outflow that has a substantial relativisitc velocity. A similar value for the Bernoulli parameter was used in e.g. \cite{Monika2014,davelaar2018}.
The second criterion is that the particle resides in high magnetisation regions where
$\sigma = B^{2}/\rho > 0.1$, where $B:=\sqrt{b_{\mu} b^{\mu}}$ is the magnetic field strength and
$b^{\mu}$ is the magnetic field 4-vector.
Satisfying this second criterion ensures that the particle ends up inside the jet sheath.
The third criterion is that the particle's radial position is at a substantial distance from the black hole,
typically $r \gtrsim 300~\rg$, at the end of the simulation.

\begin{figure}[htbp]
 \centering
 \includegraphics[width=0.9\textwidth]{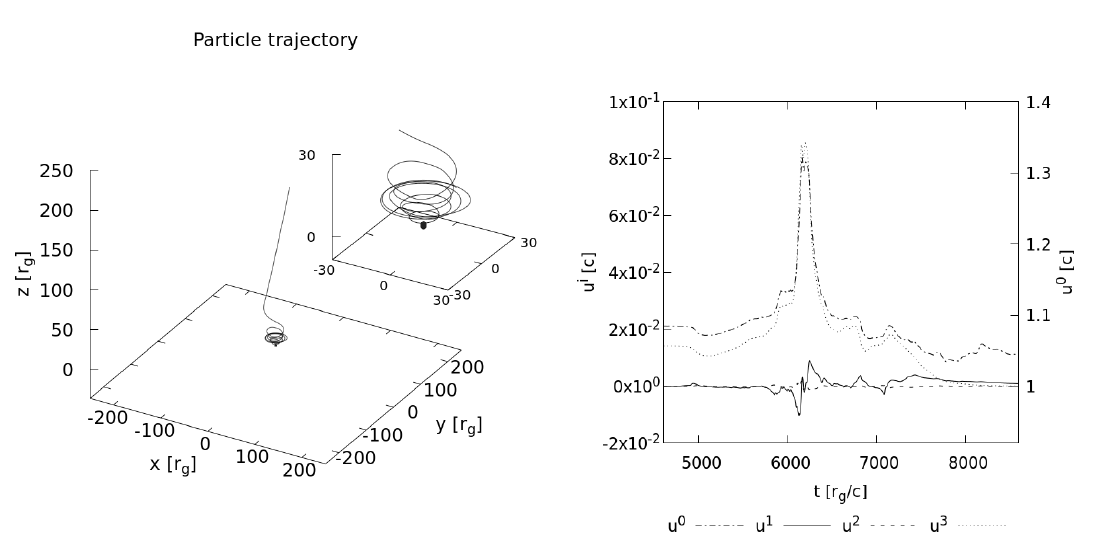}
 \caption[Used tracer-particle trajectory and it's corresponing velocity]{Left panel: the trajectory of the tracer particle that is used to initialise the camera trajectory.
  Right panel: the velocity profile of the tracer particle.
  The velocity peaks when the particle is closest to the black hole, where the angular velocity is high.
 The time shown on the x-axis is the time range of the frames used for Scene 3.}
 \label{fig:orbit}
\end{figure}

We simulate the particles with the axisymmetric GRMHD code {\tt HARM2D} \cite{HARM}.
The simulation begins with $N_{\rm tot} = 10^{5}$ particles, a simulation domain size of
$r_{\rm out} = 1000~\rg$, and is evolved until $t_{\rm final}=4000~\tunit$.
The spacetime is that of a Kerr black hole, and the dimensionless spin parameter is set to be
$a = 0.9375$.
For this value of the spin, the black hole (outer) event horizon radius is $r_{\rm h}=1.344~\rg$ and the simulation
inner boundary lies within $r_{\rm h}$ (i.e., we can track particles inside the event horizon).
The specific particle used to initialise the camera trajectory is shown in Fig.~\ref{fig:particles}
(blue square and curve).
The full particle trajectory and velocity profile for all components $u^{\rm \mu}$ are shown in
Fig.~\ref{fig:orbit}.
Rapid variations in the azimuthal 4-velocity, $u^{3}$, as well as the angular velocity, $\Omega:=u^{3}/u^{0}$,
in the right panel of Fig.~\ref{fig:orbit} are consistent with the tightly wound trajectory in the left panel.
This trajectory, which we term ``Scene 3", begins immediately after Scene 2 (i.e. after frame $4600$),
and comprises $4000$ frames, ending at frame $8599$.
\subsection{Radiative-transfer calculations and background images}

To create images of an accreting black hole, it is necessary to compute the trajectories of light rays from the radiating plasma to the observer.
For imaging applications, such as the present case, it is most computationally efficient to start the light rays at the observer instead - one for each pixel in
the image the observer sees - and then trace them backward in time. Given a ray's trajectory, the radiative-transfer equation is solved along that trajectory,
in order to compute the intensity seen by the observer. The radiative-transfer code {\tt RAPTOR} uses a fourth-order Runge-Kutta method to integrate the equations
of motion for the light rays (i.e., the geodesic equation). It simultaneously solves the radiative-transfer equation using a semi-analytic scheme (for a more detailed description of {\tt RAPTOR},
see \citet{bronzwaer2017}).
The same methodology is applied here in order to create images of the black hole accretion disk,
with one small addition.
When accretion disks, which tend to be roughly toroidal in shape, are filmed against a perfectly black background,
the resulting animations fail to convey a natural sense of motion and scale for the observer as they
orbit the black hole.
In order to increase the immersiveness of the observer and provide a physically-realistic sense of scale
and motion, the present work expands on the aforementioned radiative-transfer calculations by including an
additional source of radiation in the form of a background star field that is projected onto the celestial sphere
surrounding the black hole and observer.

This is achieved by expressing the intensity received by the observer in Lorentz-invariant form and integrating this
intensity from the camera to its point of origin within the plasma, i.e., eq.~(37) in \citep{bronzwaer2017}.
This can then be expressed in integral form (upon including a term for the background radiation) as
\begin{equation}
 \frac{I_{\nu,\rm{obs}}}{\nu_{\rm obs}^3} =
 \left(\frac{I_{\nu,\infty}}{\nu_{\infty}^3}\right) {\rm e}^{-\tau_{\nu,\rm obs}\left( \lambda_{\infty}\right)} \ +
 \int^{\lambda_{\infty}}_{\lambda_{\rm obs}}
 \left(\frac{j_{\nu}}{\nu^2}\right)
 {\rm e}^{-\tau_{\nu,\rm{obs}}}\ d{\lambda'} \,,
 \label{eqn:backward_transfer}
\end{equation}
where the optical depth along the ray is calculated as
\begin{equation}
 \tau_{\nu,\rm{obs}}\left(\lambda\right) = \int^
 {\lambda} _ {\lambda_{\rm obs}} \nu \alpha_\nu \ {\rm d}\lambda' \,.
\end{equation}
Here, $I_{\nu}$ describes a ray's specific intensity, $\nu$ its frequency, and $j_{\nu}$ and $\alpha_{\nu}$ refer
respectively to the plasma emission and absorption coefficients evaluated along the ray, which is itself
parametrised by the affine parameter, $\lambda$.
The subscript ``$\infty$" denotes quantities evaluated at the outer integration boundary
(i.e., far from the black hole), while the subscript ``${\rm obs}$'' refers to the observer's current location.
The background radiation is encoded in the term $I_{\nu,\infty}/\nu_{\infty}^3$.
The first term on the right-hand-side of eq.~\eqref{eqn:backward_transfer} is constant and represents the
intensity of the background radiation, weighted by the local optical depth.
The second term on the right-hand-side of eq.~\eqref{eqn:backward_transfer} is evaluated at a given
observer position, $\lambda_{\rm obs}$, and specifies the accumulated intensity of emitted radiation after
taking into account the local emissivity and absorptivity of the accreting plasma.
See \cite{Fuerst2004}, \cite{younsi2012}, \cite{bronzwaer2017} for further details.

A physical description of the radiation is needed for $I_{\nu,\infty} / \nu_{\infty}^3$.
Since this quantity is projected onto the celestial sphere, it is a function of two coordinates
$(\hat{\theta},\hat{\phi})$.
Note that for the ray coordinates, in the limit $r\to\infty$, both $\theta\to\hat{\theta}$ and $\phi\to\hat{\phi}$,
i.e., space-time is asymptotically flat.
We also note that only rays which exit the simulation volume (as opposed to rays which plunge
towards the horizon) are assigned a non-zero background intensity after integration.
In order to evaluate $I_\nu$ for a given ray, we therefore take the ray's
$\left(\theta,\phi\right)$ coordinates \emph{after} the ray leaves the simulation volume,
and use them as the coordinates $(\hat{\theta},\hat{\phi})$ on
the celestial sphere.
Finally, we transform these coordinates into pixel coordinates $(x,y)$ of a PNG image in order to
evaluate the intensity.
The transformation from celestial coordinates to pixel coordinates is given by
\begin{equation}
 x = \bigg\lfloor \frac{\hat{\phi}}{2 \pi} W \bigg\rfloor \quad {\rm and} \quad
 y = \bigg\lfloor \frac{\hat{\theta}}{\pi} H \bigg\rfloor \,,
\end{equation}
where $\lfloor z \rfloor \equiv {\rm floor} (z)$ is the floor function (which outputs the greatest integer $\leq z$),
and $W$ and $H$ are the width and height (in pixels) of the background image, respectively.

Using the scheme described above, it is possible to fold the background radiation field
directly into the radiative transfer calculations of the accretion disk plasma.
A second approach is to render separate movies for both the background and for the plasma,
create a composite image for all corresponding time frames between the two movies in
post-processing, and then create the new composite movie from the composite images.
We adopt the second approach in all results shown in this paper.

We have chosen a background that is obtained from real astronomical star data from the Tycho 2 catalogue
which are not in the Galactic Plane.
The original equirectangular RGB 3K image was generated by \cite{DeanVR} and converted to a
greyscale 2K image.

\subsection{Plasma and radiation models}
In this work, we seek to model the SMBH Sgr A*.
To this end we use a black hole mass of $M_{\rm BH} = 4.0\times 10^6\,\msun$ \cite{gillesen},
and a dimensionless spin parameter of $a=0.9375$, consistent with the particle simulation.
The plasma flow was simulated with the GRMHD code {\tt BHAC} \cite{BHAC}.
The simulation domain had an outer radius of $r_{\rm outer} = 1000 ~\rg$.
The simulation is initialised with a Fishbone-Moncrief torus \cite{FMtorus} with an
inner radius of $r_{\rm inner}=6~\rg$, and with a pressure maximum at $r_{\rm max} = 12~\rg$.
Magnetic fields were inserted as poloidal loops that follow iso-contours of density, and the initial magnetisation
was low, i.e., $\beta = P_{\rm gas}/B^2 = 100$, where $P_{\rm gas}$ is the gas pressure of the plasma.
The simulation was performed in three dimensions, with a resolution of 256, 128, 128 cells in
the $r$, $\theta$ and $\phi$ directions, respectively.
We simulated the flow up to $t=7000 ~\tunit$.

The GRMHD simulation only simulates the dynamically-important ions (protons).
We, therefore, require a prescription for the radiatively-important electrons in order to compute the observed emission. Most radiative models for Sgr~A* or M87 either assume that the coupling between the temperatures of the electrons and protons is constant or parameterised based on plasma variabels, see e.g. \cite{goldstone2005,noble2007,monika2009,dexter2010,Shcherbakov2012,monika2013,Monika2014,chan2015,chan2015b,gold2017}.
In this work we use, an electron model by \cite{Monika2014} where the electrons are cold
inside the accretion disk and hot inside the highly magnetized outflows.
For the electron distribution function, we adopt a thermal distribution, where \cite{davelaar2018} showed that this model accurately describes the quiescent state of Sgr~A*.
The used model \cite{Monika2014} is capable of recovering the observational parameters of Sgr A*, such as radio fluxes and
intrinsic source sizes \cite{falckemelia,bower2004,doeleman,bower2014}.

\begin{figure}[htbp]
 \centering
 \includegraphics[width=0.9\textwidth]{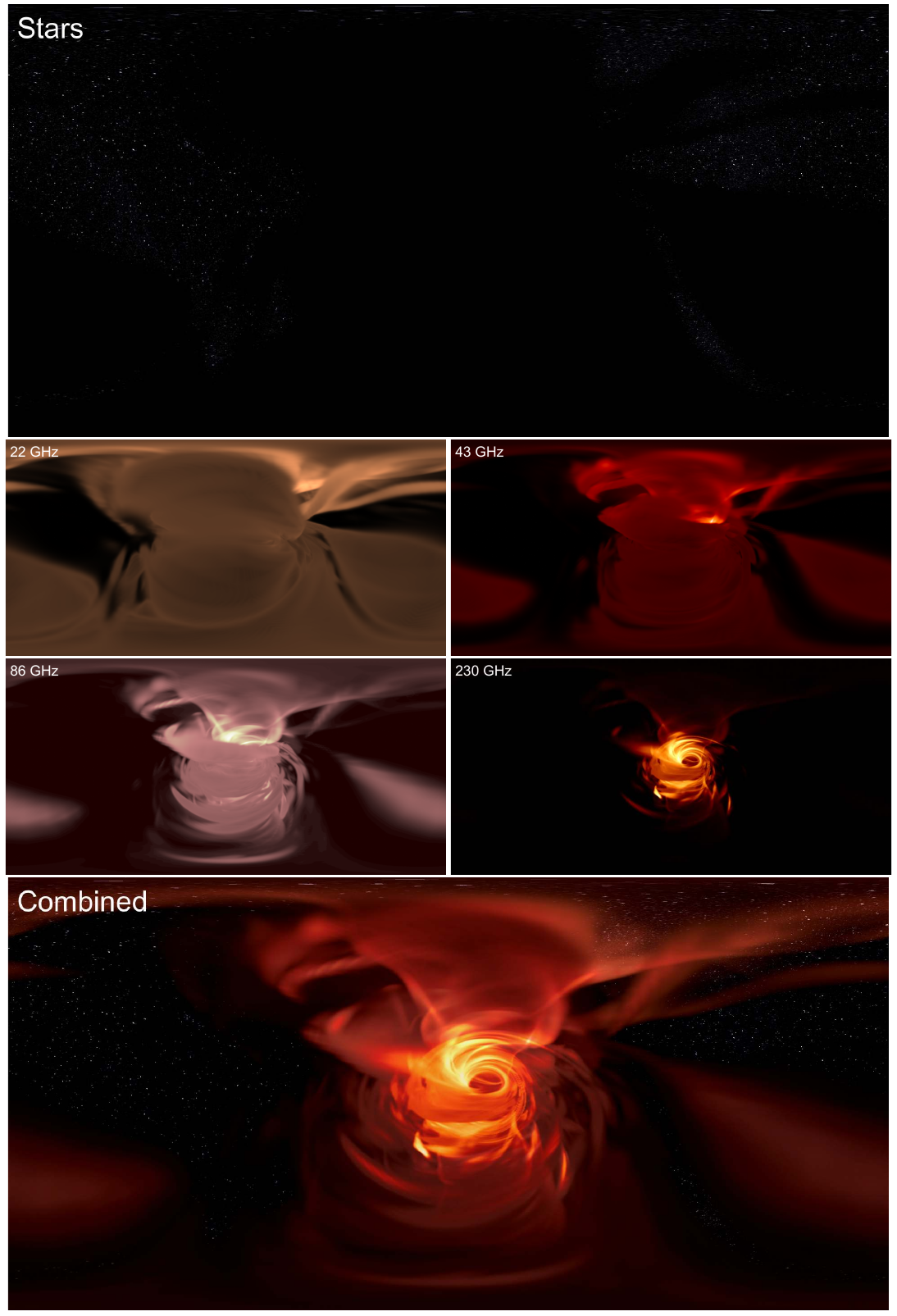}
 \caption[Images at single frequency, background and composite]{From left to right, top to bottom: snapshot panels at $t=3000~\tunit$ for:
  (i) background star field only image, (ii) $22$ GHz image, (iii) $43$ GHz image, (iv) $86$ GHz image,
  (v) $230$ GHz image, and (vi) combined (composite) image of (i)--(v).
 }
 \label{fig:averaging}
\end{figure}

We calculated the synthetic images at four different radio frequencies: $22$~GHz ($1.2$~cm),
$43$~GHz ($7$~mm), $86$~GHz ($3$~mm), and $230$~GHz ($1.3$~mm).
These frequencies were chosen since they correspond to the frequencies at which, e.g.,
the Very Long Baseline Array (VLBA) ($1.2$~mm, $7$~mm, $3$~mm), Global mm-VLBI Array (GMVA) ($3$~mm) and the Event Horizon Telescope (EHT) ($1.3$~mm) operate.
After ray-tracing these frequencies were converted into separate PNG image files, where distinct
colourmaps were chosen for each of the four frequencies.
In post-processing, these images were then combined into a single image by averaging over the RGB channels
of the four different input images.
A star-field background was also included to serve as a reference point for the observer during their motion.
This star-field background was rendered separately from the radio images,
although the opacity at $22$ GHz was used to obscure stars located behind the accretion disk.
This background was then also averaged together with the radio images using the same RGB channel averaging.
The four separate frequencies, the star-field background, and the resulting combined image are
presented in Fig.~\ref{fig:averaging}.
\section{\label{sec:results} VR movie}
The resulting VR movie contains $8600$ frames at a resolution of $2000\times1000$ pixels. As a proof of concept, this resolution was chosen to balance image quality and computational resources. Current VR headsets also upscale the provided resolution with interpolation routines. We tested the resolution with the Oculus VR headset, which turned out to be sufficient. Since the provided methods are not limited by the resolution, a larger resolution can in principle be achieved. The movie is available on Youtube VR \cite{MovieVR}.
In this section, we discuss several snapshots from this movie.

\begin{figure}[htbp]
 \centering
 \includegraphics[width=0.9\textwidth]{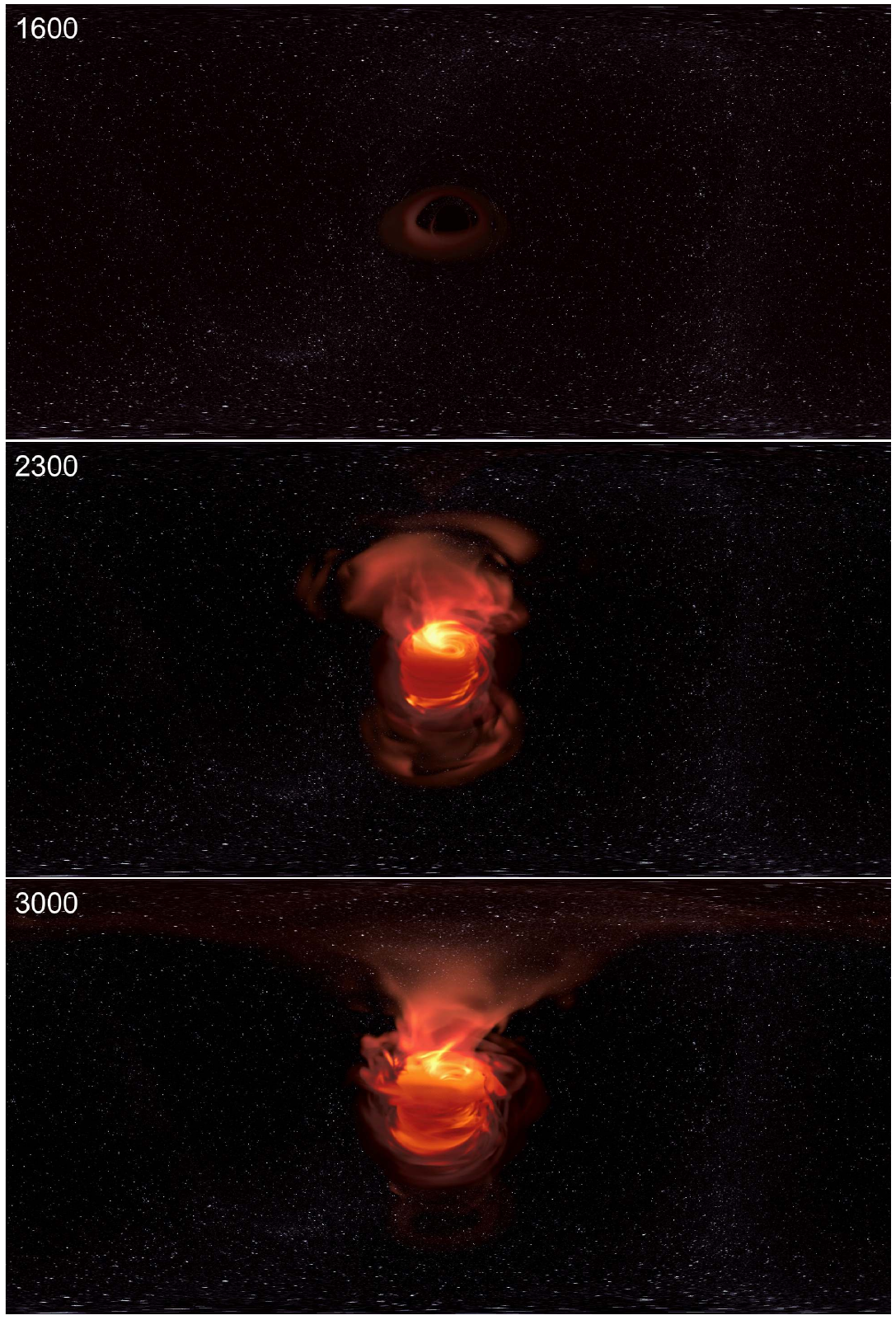}
 \caption[Various snapshots of Scene 1]{Movie snapshots from Scene 1.
  The simulation time (in units of $\tunit$) is shown in the upper-left corner of all panels.
  From top to bottom:
  Scene 1 begins at frame $1600$, where accretion onto the black hole has not yet begun, which can be seen
  as the faint, stationary equilibrium accretion torus configuration in the centre of the image.
  By frame $2300$ accretion has begun (see also Fig.~\ref{fig:accrate}) and the dim jet (upper half of image)
  and dimmer counter jet (lower half of image) propagate outwards through the ambient medium.
  At frame $3000$ the jet has propagated further outwards, and angular momentum transport has shifted
  torus material outward, as can be seen by the increased angular size of the inner accretion flow.
  The black hole shadow is not visible since the accretion rate has yet to reach a quasi-stationary state.
 }
 \label{fig:snapshots-scene1}
\end{figure}

\begin{figure}[htbp]
 \centering
 \includegraphics[width=0.9\textwidth]{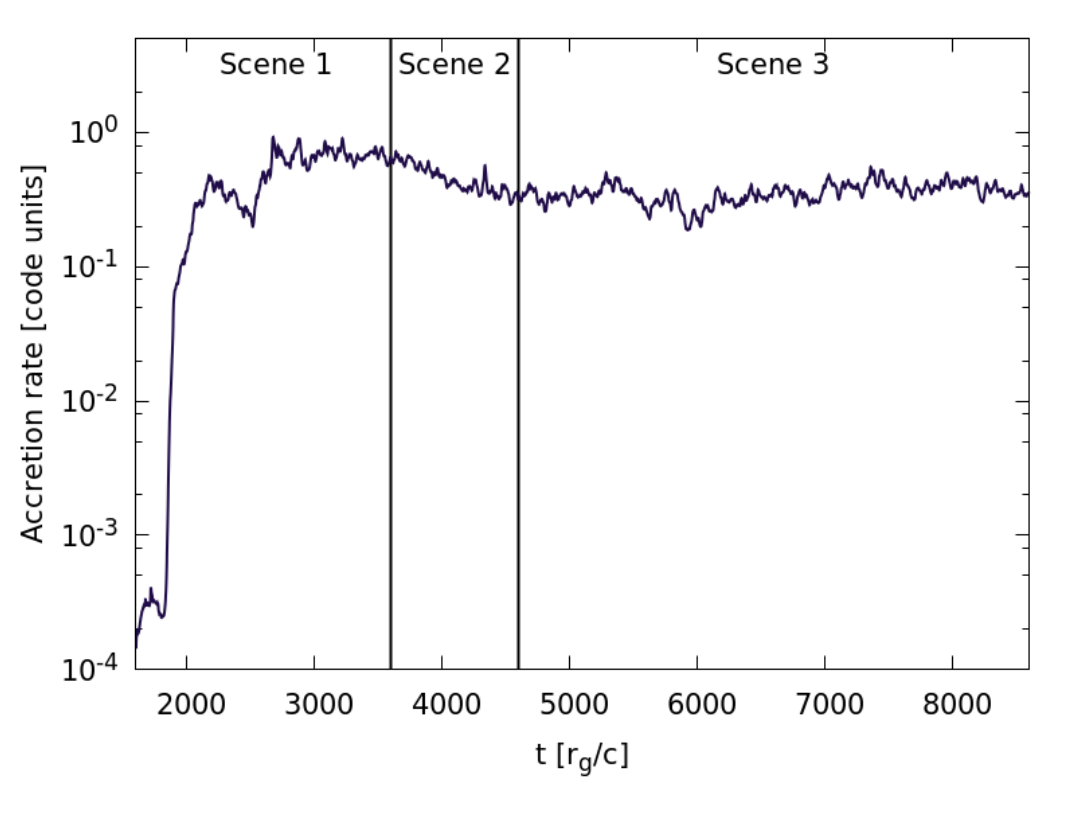}
 \caption[Accretion rate as a function of time]{Simulation accretion rate as a function of time (in code units). At t=2500 the MRI start to saturate. The time shown on the x-axis is the time of the frames used for ``Scene 2'' and ``Scene 3''.
 }
 \label{fig:accrate}
\end{figure}

The first set of snapshots is shown in \ref{fig:snapshots-scene1}.
In Fig.~\ref{fig:snapshots-scene1} we show a set of snapshots from Scene 1, ($1600$, $2300$, $3000$),
matter starts to accrete onto the black hole and the jet is launched.
The jet then propagates through the ambient medium of the simulation, forming a collimated funnel that is mainly
visible at lower frequencies.
Since the accretion rate peaks at this point in the simulation (see Fig.~\ref{fig:accrate}),
the black-hole shadow is barely visible.

\begin{figure}[htbp]
 \centering
 \includegraphics[width=0.9\textwidth]{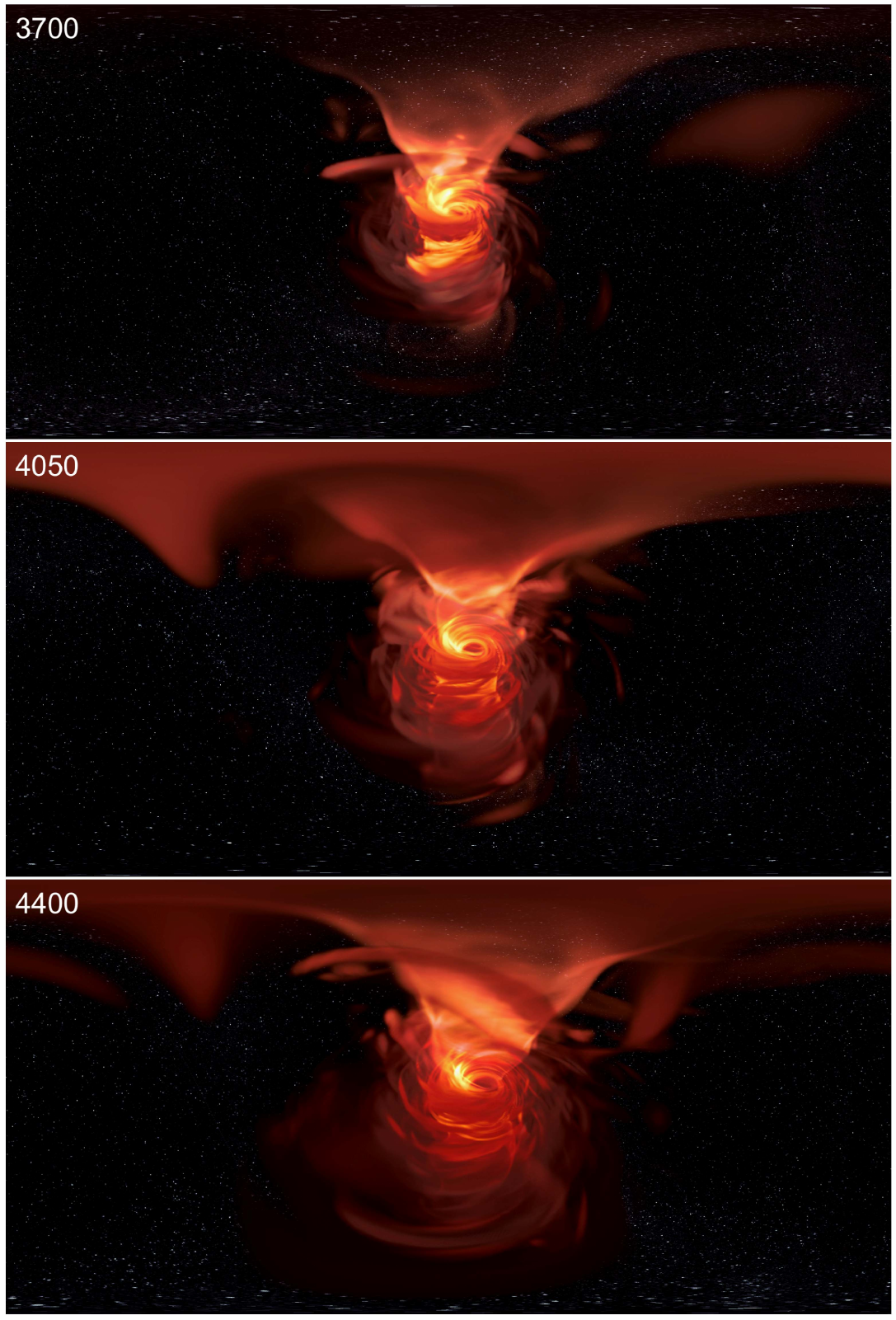}
 \caption[Various snapshots of Scene 2]{Movie snapshots from Scene 2.
  By frame $3700$ the MRI has begun to saturate and the accretion rate reaches a quasi-stationary state.
  At frame $4050$ the jet and counter-jet have propagated further away from the black hole and reached the boundary
  of our simulation domain.
  Due to the steadier accretion rate, by frame $4400$ the central region surrounding the event horizon becomes cooler
  and more optically thin.
  The upper-half of the black-hole shadow is now visible.
 }
 \label{fig:snapshots-scene2}
\end{figure}

In Fig.~\ref{fig:snapshots-scene2} we show snapshots from Scene 2 ($3700$, $4050$, $4400$),
the jet propagates outward to the boundary of our simulation domain, the accretion rate settles
and the black hole shadow becomes visible.

\begin{figure}[htbp]
 \centering
 \includegraphics[width=0.9\textwidth]{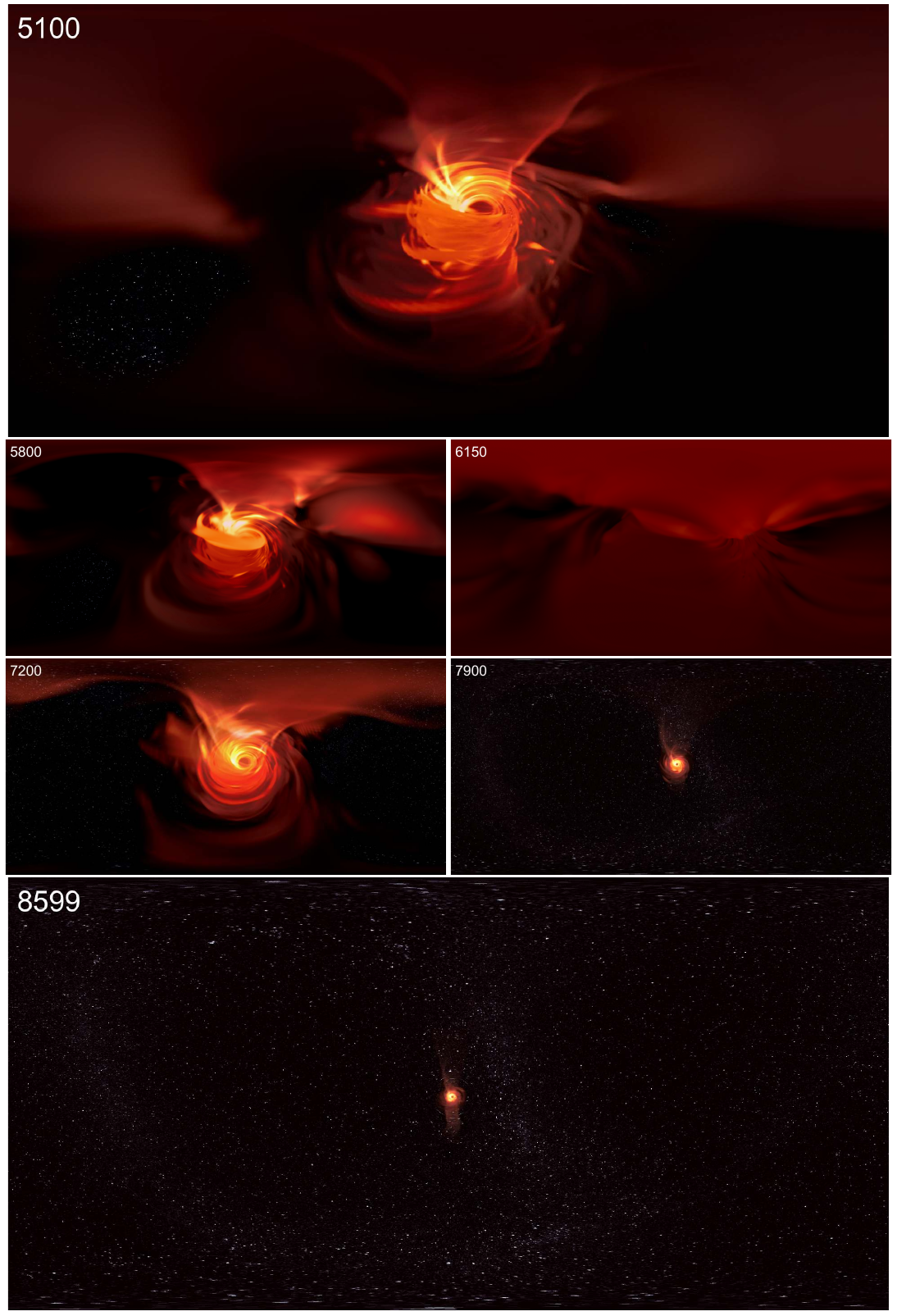}
 \caption[Various snapshots of Scene 3]{Movie snapshots from Scene 3.
  The observer now begins their journey through the accretion flow (panels with frames $5100$--$6150$),
  before being advected away from the black hole via the large-scale jet (panels with frames $7200$--$8599$).
  At frame $6150$ the observer is at their point of closest approach to the black hole, where the incident flux is
  as high as $\approx 25 L_{\odot}$.
  This region is highly optically thick, completely obscuring the observer's view of the black hole shadow.
  As the observer is advected further away, by frame $8599$ the angular size of the black hole and the surrounding
  accretion flow is greatly reduced and appears almost point-like.
 }
 \label{fig:snapshots-scene3}
\end{figure}

In Fig. \ref{fig:snapshots-scene3} we show snapshot from Scene 3.
When the observer moves along with the flow in Scene 3 ($5100$, $5800$, $6150$), small hot blobs of plasma
orbiting the black hole are distinguishable.
At closest approach (around $6~\rg$, frame $6150$), the scene changes rapidly.
This is due not only to rapid rotation of the black hole but also to the rapid decrease of observed flux.
It is hard to distinguish individual stars and the only observable emission is at $230$ GHz.
At the end of Scene 3 ($7200$, $7900$, $8599$) the observer exits the accretion disk via the jet, whereafter
a rapid increase in radial velocity is clearly seen.

\begin{figure}[htbp]
 \centering
 \includegraphics[width=0.9\textwidth]{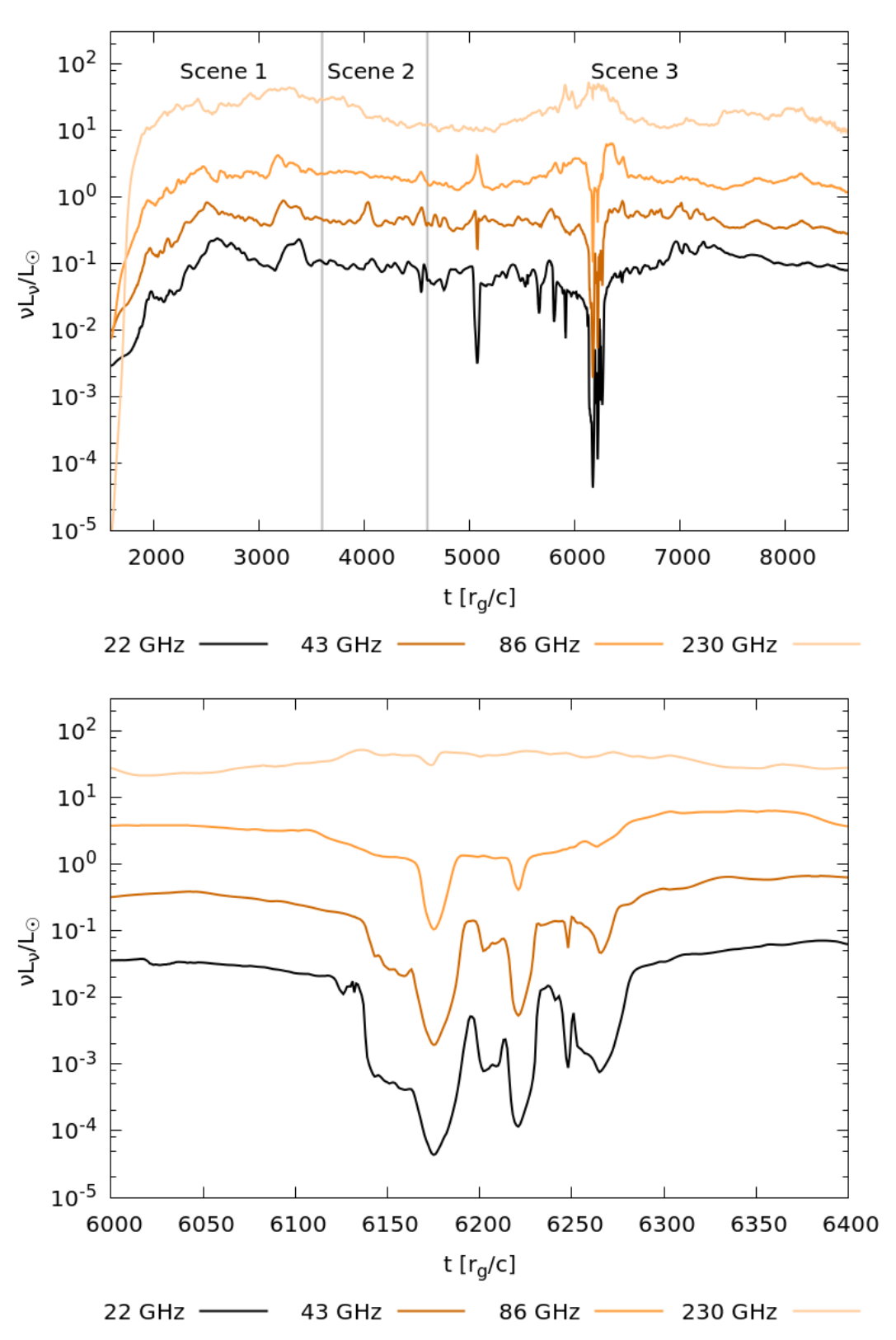}
 \caption[Luminosity as a function of time]{Top panel: total luminosity collected at the camera at each time step.
  Bottom panel: magnified view of the time range $6000$--$6400~\tunit$, where the camera passes
  through the optically thick part of the accreting plasma.
 }
 \label{fig:lightcurves}
\end{figure}

To obtain a better quantitative understanding of the movie we also calculate the total bolometric luminosity as
received by the observer's camera.
This is shown in the top panel of Fig.~\ref{fig:lightcurves}.
At $6150$ a decrease in luminosity is evident at the three lowest frequencies, which corresponds to where the
observer is closest to the black hole event horizon and has entered the optically-thick accretion disk.
A magnified version of this Figure in the optically-thick part is shown in the bottom panel of Fig.~\ref{fig:lightcurves}.
A frame corresponding to this particular moment is shown in Fig.~\ref{fig:snapshots-scene3}, panel $6150$.
At closest approach, the total luminosity detected at $230$ GHz peaks, and the observer is exposed to
$\approx 25 L_\odot$.

\section{\label{sec:discconlc} Discussion and conclusion}
In this work, we have detailed our methods for visualising the surroundings of accreting black holes in virtual reality.
We presented a visualisation of a three-dimensional fully-general-relativistic accreting black hole simulation
in a full $360^{\circ}$ VR movie with radiative models based on physically-realistic GRMHD plasma simulations.
In order to produce representative images, the radiative-transfer capabilities of our code {\tt RAPTOR} were
extended to include background starlight and an observer in an arbitrary state of motion.
To model the emission emerging from the vicinity of a black hole we coupled the GRMHD simulation with our
radiative-transfer code to produce a VR movie based on our recent models for Sgr~A*
\cite{Monika2014,davelaar2018}.
These methods can be applied to accreting black holes of any size, so long as radiation feedback onto
the accretion flow has a negligible impact on the flow's magnetohydrodynamical properties.

The trajectory of the camera consisted of two phases: a hovering observer and an advected observer.
For this second phase, we used an axisymmetric GRMHD simulation, in contrast to the plasma simulation used
to calculate the radiation, which was fully-three-dimensional.
This choice, whilst scientifically less accurate, was intentional and somewhat necessary.
Turbulent features in the $\phi$ direction were omitted since they can be nauseating
to watch in VR environments and commonly lead to motion sickness.
A composition of starfield and accretion flow images at four frequencies was then used to create a movie,
consisting of $8600$ frames, which is freely available on YouTube.

This movie couples GRMHD simulations with GRRT post-processing in VR. Since we do not make any strong a-priori assumptions regarding the field-of-view of the observer,
we can calculate the full radiation field measured at a specific point in the accretion disk,
where we include all GR effects.
This enabled us to calculate light curves of the total measured luminosity at multiple frequency bands at the
position of a particle being advected in the flow.
This way of calculating the full self-irradiation of the disk is of potential interest in, e.g.,
studies of X-ray reflection models in AGN, or coupling to GRMHD simulation to calculate the proper
radiative feedback onto an emitting, absorbing (and even scattering) plasma in GR in a self-consistent way.

Finally, beyond the aforementioned scientific applications, VR represents a new medium for
scientific visualisation which can be used, as demonstrated in this work, to investigate the emission that an observer would measure from \emph{inside} the accretion flow.
It is natural, and of contemporary interest even in the film industry
\citep[see e.g.][]{James2015a,James2015b}
to ask the question as to what an observer would see if they were in the immediate vicinity of a black hole.
In this work, we have sought to address this question directly, by using state-of-the-art numerical techniques and
astrophysical models in a physically-self-consistent manner.
Given the EHTC is anticipated to obtain images of the black hole shadows in Sgr~A* and M87 in the near future,
the calculations we have presented are timely.
The VR movies presented in this work also provide an intuitive and interactive way to communicate black hole
physics to wider audiences, serving as a useful educational tool.

\begin{backmatter}
 
 \section*{Abbreviations}
 
 Alphabetical list of the abbrevitions used in the main text;
 
 \begin{enumerate}[label={}]
  \item {AGN -} Active Galactic nuclei
        
        \item{CPUs -} Central Processor Units
        
        \item{EHT -} Event Horizon Telescope
        
        \item{EHTC -} Event Horizon Telescope Collaboration
        
        \item{GMVA - } Global mm-VLBI Array
        
        \item{GPUs -} Graphical Processor Units
        
        \item{GR -} Einstein's General Theory of Relativity
        
        \item{GRMHD -} General Relativistic MagnetoHydroDynamics
        
        \item{GRRT -} General-Relativistic Radiative-Transfer
        
        \item{MKS -} Modified Kerr-Schild
        
        \item{Sgr~A* -} Sagittarius ~A*
        
        \item{SMBH -} SuperMassive Black Hole
        
        \item{VLBA -} Very Long Baseline Array
        
        \item{VLBI -} Very Long Baseline Interferometry

        \item{VR -} Virtual Reality
 \end{enumerate}
 
 \section*{Availability of Data and Materials}
 Please contact author for data requests.
 
 \section*{Competing interests}
 The authors declare that they have no competing interests.
 
 \section*{Funding}
 The authors acknowledge support from the ERC Synergy Grant ``BlackHoleCam: Imaging the Event Horizon of Black Holes" (Grant 610058). ZY acknowledges support from a Leverhulme Trust Early Career Fellowship.
 
 \section*{Author’s contributions}
 JD performed GRMHD and GRRT simulations, created the movie, and wrote the initial manuscript. TB provided the {\tt RAPTOR} code. JD and TB extended {\tt RAPTOR} to generate $360^{\circ}$ images. DK performed the particle-tracer simulation. TB and ZY helped write the manuscript. ZY provided code to calculate background lensing
 structures. JD, MM, TB, and HF provided the physically motivated model for
 Sgr A*. MM provided initial guidance on how to make $360^{\circ}$ movies. ZY, MM and HF provided ideas to initialise the project and provided feedback throughout. All authors discussed and commented on the final manuscript.
 
 \section*{Acknowledgements}
 The authors thank Oliver Porth, Sera Markoff, Dimitrios Psaltis, Chi-kwam Chan, Christiaan Brinkerink, Yosuke Mizuno, Luciano Rezzolla, and Robin Sip, for useful comments and discussions during this project. The GRMHD simulation was performed on the {\tt LOEWE} computing facility at the CSC-Frankfurt, the advection simulation and radiative-transfer calculations were performed on the {\tt COMA} computing facility at Radboud University Nijmegen. This research has made use of NASA's Astrophysics Data System.

 \nocite{*}
 
 \bibliographystyle{bmc-mathphys}

 \listoffigures
 
\end{backmatter}
\end{document}